# Crystal structure, magnetic and resonant properties of decorated spin kagome system (CsCl)Cu$_5$As$_2$O$_{10}$

Ilya V. Kornyakov,*[a,b] Marina V. Likholetova,[a] Irina E. Lezova,[a] Sergey V. Krivovichev,[a,b] Harald O. Jeschke,[c,d] Yasir Iqbal,[d] Alexey V. Tkachev,[e] Sergey V. Zhurenko,[e,f] Andrey A. Gippius,[e,f] Larisa V. Shvanskaya*[f,g] and Alexander N. Vasiliev[f,g]

We report the synthesis and investigation of crystal structure and physical properties of the averievite-like arsenate, (CsCl)Cu$_5$As$_2$O$_{10}$. Just above room temperature, this compound undergoes a structural phase transition from the high temperature trigonal $P\bar{3}m1$ ($a = b = a_0$ and $c = c_0$) to the low temperature monoclinic $I2/a$ phase ($a \approx \sqrt{3}a_0$, $b = a_0$, $c \approx 2c_0$ and $\beta \approx 90.6°$). According to thermodynamic and nuclear magnetic resonance measurements, it experiences a phase transition into a canted antiferromagnetic state at $T_N$ = 21 K. The density functional theory calculations place the energy scale of kagome exchange interaction parameter in (CsCl)Cu$_5$As$_2$O$_{10}$ in between those in V- and P-analogs.

## Introduction

Geometrically frustrated layered systems are considered as a potential platform for realizing the quantum spin liquid state in solids.[1] The kagomé lattice, which is a two-dimensional arrangement of regular triangles and hexagons belongs to these systems.[2] In recent decades, this arrangement of magnetically active ions has been found in the structure of copper-rich minerals, herbertsmithite, ZnCu$_3$(OH)$_6$Cl$_2$,[3] volborthite, Cu$_3$V$_2$O$_7$(OH)$_2$·2H$_2$O,[4] vesignieite, BaCu$_3$V$_2$O$_8$(OH)$_2$,[5] francisite, Cu$_3$Bi(SeO$_3$)$_2$O$_2$Cl,[6] ilinskite, (NaCu$_5$O$_2$(SeO$_3$)$_2$Cl),[7] avdoninite, K$_2$Cu$_5$Cl$_8$(OH)$_4$·2H$_2$O,[8,9] and averievite, (CsCl)Cu$_5$V$_2$O$_{10}$.[10] The magnetic subsystem of the last three species features additional magnetically active ions located both above and below the kagomé lattice triangles, completing them into tetrahedra. This modification is considered also as pyrochlore slab and leads to additional magnetic exchanges within the Cu-based tetrahedra.

Recently, a series of averievite-like compounds, (MX)Cu$_5$T$_2$O$_{10}$ was synthesized with T = P, V and M = K, Rb, Cs, Cu$^+$ and X = Cs, Br.[11-14] The successful growth of single crystals allowed establishing the cascade of phase transitions and large magnetic anisotropy for the phosphate analog of averievite, (CsCl)Cu$_5$P$_2$O$_{10}$.[15] It undergoes a first-order structural transition at around 224 K and two successive antiferromagnetic phase transitions at 13.6 K and 2.2 K.

Averievite, (CsCl)Cu$_5$V$_2$O$_{10}$, and its phosphate analog, (CsCl)Cu$_5$P$_2$O$_{10}$, have been studied in density functional theory (DFT) calculations,[16] and exchange interactions up to fourth nearest neighbors have been worked out. The model Hamiltonian of the pyrochlore slab, which was termed capped kagomé has been studied by a density matrix renormalization group technique.[17] The properties of the classical pyrochlore slab magnet have also been investigated.[18]

Synthetic averievite, (CsCl)Cu$_5$V$_2$O$_{10}$, possesses a paramagnetic Curie-Weiss temperature Θ = − 185 K and orders antiferromagnetically at $T_N$ = 24 K. It has been shown that the long-range order is suppressed upon substitution of copper by zinc.[11] The strength of the exchange interactions can be influenced also by varying the cation in tetrahedral $T$O$_4$ groups ($T$ = P, V). Since the arsenates can be isostructural to phosphates and vanadates, we undertook the synthesis and investigation of crystal structure and physical properties of the averievite-like arsenate, (CsCl)Cu$_5$As$_2$O$_{10}$.

## Experimental

### Synthesis

The synthesis of (CsCl)Cu$_5$As$_2$O$_{10}$ has been done in two steps, i.e., the hydrothermal synthesis of the precursor Cu$_3$(AsO$_4$)$_2$, followed by the solid-state synthesis of the title compound. The details of synthesis of Cu$_3$(AsO$_4$)$_2$ are given in Supporting Information (SI).

The powder sample of (CsCl)Cu$_5$As$_2$O$_{10}$ was synthesized via solid-state reactions in air. Three reagents, CuO (LenReactiv, 99.9 %), CsCl (AlfaAesar; 99.999 %) and pre-obtained Cu$_3$(AsO$_4$)$_2$, were mixed in a molar ratio of 2:1:1 and ground in an agate mortar. The resulting mixture was loaded into a fused

[a.] St. Petersburg State University, St. Petersburg 199034, Russia. E-mail: i.kornyakov@spbu.ru
[b.] Kola Science Centre, RAS, Apatity 184209, Russia.
[c.] Okayama University, Okayama 700-8530, Japan.
[d.] Indian Institute of Technology Madras, Chennai 600036, India.
[e.] Lebedev Physical Institute, RAS, Moscow, Russia.
[f.] Moscow State University, Moscow 119991, Russia. E-mail: lshvanskaya@mail.ru
[g.] National University of Science and Technology MISIS, Moscow 119049, Russia.
Supplementary Information available: additional crystallographic data obtained via X-Ray diffraction and details of the DFT energy mapping. See DOI: 10.1039/x0xx000

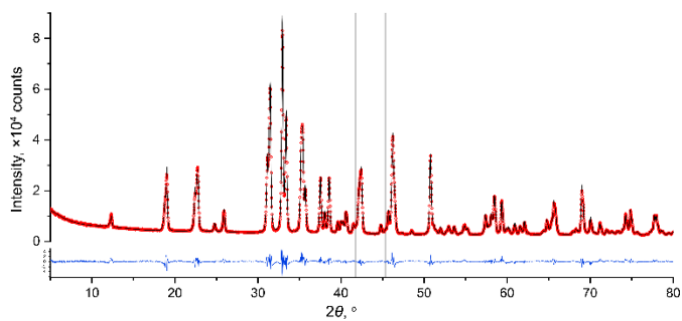

**Figure 1** Rietveld fit (black line) of the powder X-ray diffraction pattern of $(CsCl)Cu_5As_2O_{10}$ collected at 295 K (red points). The gray lines indicate impurity reflections of CuO.

quartz ampoule (~5 cm in length, ~0.4 cm inner diameter, 0.2 cm wall thickness) and compressed into a pellet using a carbon rod. The ampoule was positioned vertically in a muffle furnace and heated to 640 °C over 4 hours. After 28 hours at this temperature, the furnace temperature was decreased to 525 °C over 2 hours and held constant for 8 hours. The final step of the synthesis involved cooling to 400°C over 36 hours, after which the furnace was switched off. After cooling to room temperature, the product was a yellowish-green powder. Powder X-ray diffraction analysis showed that the product primarily consisted of $(CsCl)Cu_5As_2O_{10}$ (Fig. 1), with a minor admixture of CuO (<0.7 %). All attempts to obtain a nominally pure sample were unsuccessful.

Single crystals of $(CsCl)Cu_5As_2O_{10}$ were obtained using a similar methodology to that described for the powder sample. A mixture with the same composition was placed in a quartz ampoule (~16 cm in length, ~0.8 cm inner diameter, 0.1 cm wall thickness), which was evacuated to a pressure of approximately $10^{-2}$ mbar and sealed. The ampoule was placed horizontally in a furnace and heated to 700 °C over 4 hours, then held at this temperature. After 200 hours, the temperature was reduced to 500°C over the course of 300 hours, after which the furnace was switched off. Upon cooling, the ampoule contained hexagonal single crystals of $(CsCl)Cu_5As_2O_{10}$ up to 0.3 mm in size, red crystals of $Cu_2O$ and blue crystals of the initial $Cu_3(AsO_4)_2$.

**Powder X-ray diffraction**

The phase composition of the obtained polycrystalline sample was analyzed by powder X-ray diffraction (PXRD) using a Rigaku Miniflex II diffractometer equipped with a $CoK\alpha_{1+2}$ X-ray tube ($\lambda(CoK\alpha_1)$ = 1.78896 Å; $\lambda(CoK\alpha_2)$ = 1.79285 Å). Diffractograms were recorded over a $2\vartheta$ range of 5–80° with a step size of 0.02° and a scan speed of 2°/min. The sample was rotated at a speed of 20 rpm during data acquisition. Phase identification was performed using PDXL 2 software[19] with the ICDD's PDF-2 (release 2020) powder database.[20] The presence of each phase was confirmed by fitting its structural data, imported from ICSD (ver. 4.8.0), using the Whole Powder Pattern Fitting method in the TOPAS 5 software.[21,22]

Initial screening revealed that the crystal undergoes a phase transition from trigonal to monoclinic symmetry upon cooling, accompanied by twinning. To account for this, two data collection strategies were devised, employing different exposure times (2 and 5 seconds) and detector-to-sample distances (34 and 50 mm) for the temperature ranges 400–300 K and 300–100 K, respectively. The data collection strategy for the trigonal phase was designed to acquire a hemisphere with a redundancy of 5, a frame width of 0.5°, and a $2\vartheta$ angular limit of 84°. For the monoclinic phase, the strategy was adjusted with a $2\vartheta$ angular limit of 76°.

*In situ*, thermal behavior of $(CsCl)Cu_5As_2O_{10}$ was investigated using a Rigaku Ultima IV powder X-ray diffractometer (PXRD, CuK$\alpha$ radiation; 40 kV/30 mA; Bragg-Brentano geometry; PSD D-Tex Ultra detector). A Rigaku R-300 chamber was used for data collection in vacuum within a range of 100–400 K on cooling and on heating cycles consequently, with the temperature steps set at 20 K (for the 320–400 and 100–280 K ranges) and 4 K (for the 280–320 K range). The heating rate was 2 K min$^{-1}$. Diffractograms were recorded over a $2\vartheta$ range of 7–70° with a step size of 0.02° and a scan speed of 4°/min. A powder sample as a heptane suspension was placed on a Cu strip (20 × 12 × 1.5 mm$^3$). The zero-shift parameter was refined at every step, and it was usually increased by 0.01–0.02° $2\vartheta$ because of the sample holder expansion upon heating. The unit-cell parameters were calculated at every temperature step by the method described in Refs. [21, 22]. The final $R_{exp}$, $R_{wp}$ and GoF parameters are in the ranges of 1.87–1.88, 2.26–2.8 and 1.21–1.46, respectively.

**Single-crystal X-ray diffraction**

A single crystal of $(CsCl)Cu_5As_2O_{10}$ was selected under an optical microscope and mounted on the glass fiber using epoxy. X-ray diffraction data were collected using a Rigaku XtaLAB Synergy-S X-ray diffractometer equipped with a monochromatic microfocus MoK$\alpha$ PhotonJet-S source ($\lambda$ = 0.71073 Å; 50 kV, 1.0 mA) and a HyPix 6000HE hybrid photon counting detector. A temperature-dependent single-crystal X-ray diffraction (SCXRD) experiment was conducted using an Oxford Cobra cooling system. Temperature steps were set to 10 K (300–400 K), 3 K (290–296 K), and 20 K (100–280 K) during cooling.

CrysAlisPro software[23] was used for the integration and correction of diffraction data for polarization, background and Lorentz effects. Absorption corrections were applied using a numerical Gaussian integration over a multifaceted crystal model and an empirical method based on spherical harmonics *via* the SCALE3 ABSPACK algorithm. Manual inspection of the data revealed the first signs of a phase transition at 290 K, characterized by the appearance of additional reflections associated with the monoclinic cell (Fig. S1 of Supporting Information (SI)). However, these reflections were of low intensity, preventing successful data processing in a monoclinic setting. Consequently, diffraction data at 293 K were processed as trigonal. Below 290 K, all data were processed as monoclinic in an *I*-cell. Twinning was addressed using CrysAlisPro, which allowed separation of diffraction data for each twin domain. The final dataset included HKLF4- and HKLF5-type reflection files, containing reflections from the primary domain and all associated overlapped reflections.

The unit-cell parameters were refined using a least-squares technique. The structure was solved by a dual-space algorithm and refined using SHELX programs,[24,25] which were incorporated in the OLEX2 program package.[26]

## Results and discussion

### Crystallography

Fig. 2, as well as Fig. S2 and tables S1 and S2 of the SI present the changes in the unit-cell parameters of $(CsCl)Cu_5As_2O_{10}$ upon cooling, as determined by powder and single-crystal X-ray diffraction. Within the temperature range of 296–304 K (according to PXRD), the compound undergoes a structural phase transition from the trigonal $P\bar{3}m1$ ($a = b = a_0$ and $c = c_0$) to the $I2/a$ monoclinic phase ($a ≈ \sqrt{3}a_0$, $b = a_0$, $c ≈ 2c_0$ and $β ≈$ 90.6°).

Although the monoclinic cell has been previously observed in several averievite-type phases,[12,14,27] the monoclinic cell observed in $(CsCl)Cu_5As_2O_{10}$ is distinct due to its $I$-centering. PXRD data indicate that in the 296–304 K range, both trigonal and monoclinic phases coexist, which is not uncommon for structural phase transitions. However, as evident from Fig. S1 of the SI, the coexistence of both phases is also observed in the single crystal. Notably, at 290 K, additional reflections corresponding to the monoclinic phase become visible, while the main reflections from the trigonal phase remain unsplit – a characteristic feature of a structural phase transition.

Fig. 2 and Fig. S3 of the SI illustrate the temperature-dependent changes in the unit-cell parameters of $(CsCl)Cu_5As_2O_{10}$. Upon cooling, the unit-cell parameters of the trigonal phase gradually decrease until 304 K, where the structural phase transition begins, accompanied by the coexistence of both trigonal and monoclinic phases. By 292 K, only the monoclinic phase persists. As the temperature decreases further, the degree of "monoclinicity" increases, as evidenced by the rise in the $β$ unit-cell parameter (Fig. S3c of the SI). Concurrently, the $c$ and $b$ parameters decrease, leading to a reduction in the unit-cell volume (Fig. S3a,b,d of SI). Meanwhile, the $a$ parameter remains relatively constant

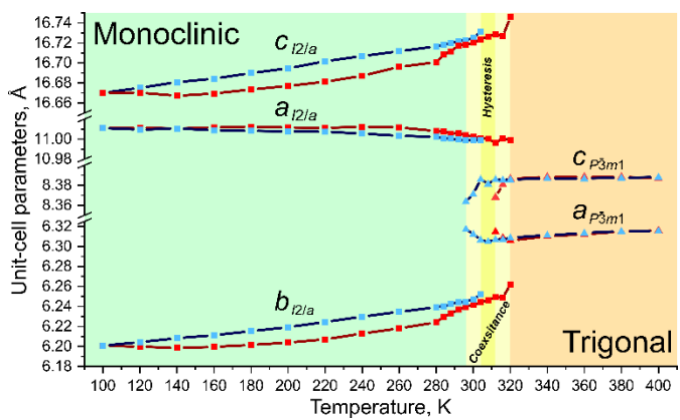

**Figure 2** Changes in the unit-cell parameters of $(CsCl)Cu_5As_2O_{10}$ upon cooling (blue symbols) and heating (red symbols). Legend: triangles - trigonal phase; squares - monoclinic phase.

throughout the process (Fig. S3a of SI).

Upon heating the unit-cell volume continues to reduce. Only within the range of 140–160 K do the unit-cell parameters start to increase, with the exception of the $β$ angle, which maintains the trend observed upon cooling up to ~220 K. At 280 K, the trend for cell expansion changes significantly, characterized by a change in the sign of a second-order polynomial fit (Table S3 of SI). The $b$ parameter is the only one to consistently decrease, starting from 240 K. The unit cell then rapidly expands up to 312 K, at which point the trigonal phase begins to emerge. Between 316 K and 320 K, the monoclinic cell undergoes an even more pronounced expansion ($ΔV ≈ 3.4$ Å$^3$) until the complete vanishing of the phase at 340 K. Beyond this temperature, the unit-cell parameters of the trigonal phase increase gradually, with a rate approximately matching that observed during cooling.

The crystal structure of averievite-type compounds is well-documented and has been extensively described in previous studies.[10-15, 27-29] The key structural motif is the capped-kagomé layer or the [111] pyrochlore slab (Fig. 3a), which comprises corner-sharing, anion-centered $OCu_4$ tetrahedra arranged in a **UDUDUD** pattern (where **U** and **D** indicate that the non-bonded vertices of the tetrahedra point **U**p and **D**own relative to the layer plane). These layers are further decorated by ($T^{5+}O_4$)$^{3-}$ tetrahedra ($T^{5+}$ = V, As, P) which are attached to the bases of the anion-centered tetrahedra in a 'face-to-face' configuration. The remaining vertices of the ($T^{5+}O_4$)$^{3-}$ tetrahedra are directed toward the vertices of the anion-centered tetrahedra in adjacent layers, forming an electroneutral framework $\{[O_2Cu_5](T^{5+}O_4)_2\}^0$ (Fig. 3b). Due to the kagomé layer topology, the framework contains channels that are occupied by $M^+$ and $X^-$ ions ($M^+$ = Cs, Rb, K, Cu; $X^-$ = I, Br, Cl). The $X^-$ anions are positioned within the kagomé layer plane, while the $M^+$ cations reside between the layers.

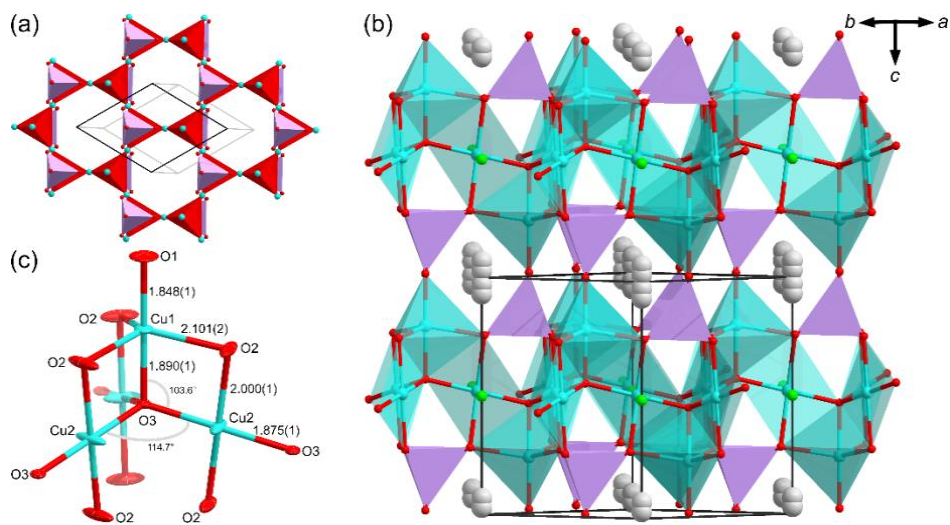

**Figure 3** Pyrochlore slab of $OCu_4$ tetrahedra (a), crystal structure (b) and $Cu^{2+}$-centered coordination geometries (c) of trigonal $(CsCl)Cu_5As_2O_{10}$ at 290 K. Bond distances in (c) are averaged over the entire temperature range. Legend: Cs - gray; Cu - cyan; As - purple; O - red; Cl - green.

In the archetypal trigonal setting, a single anion-centered tetrahedron is formed by two symmetrically distinct $Cu^{2+}$ cations with triangular-bipyramidal (Cu1) and square-planar (Cu2) coordination geometries (Fig. 3c). Both geometries remain stable upon cooling, showing no significant changes in bond lengths (Table S4 of SI). The only notable feature is the increase of $U_{22}$ anisotropic displacement parameter of the Cu2 atom (and coordinating O2 atom) from 0.062 to 0.075 Å$^2$.

The structural phase transition induces a significant reconstruction of the structural framework, primarily manifesting in the rearrangement of oxygen ligands around the Cu2 atom. The distortion of the Cu1O$_5$ triangular bipyramid arises from the elongation of one equatorial bond (Cu1–O2) from ~2.10 Å to 2.33 Å, resulting in its transformation into a distorted square pyramid (Fig. 4a). Furthermore, upon cooling, this bond continues to elongate, reaching ~2.44 Å. In contrast, the remaining bonds within the Cu1O$_5$ polyhedron exhibit much less pronounced changes: bonds with additional oxygen atoms remain largely unchanged, while those with O$_{As}$ atoms gradually shorten (Table S5 of SI). The bond length variations in the $Cu^{2+}$-centered square planes are minor and can be characterized as fluctuations without significant structural impact.

Along with distortions in the first coordination spheres of $Cu^{2+}$-centered polyhedra, the kagomé layer undergoes corresponding distortions, primarily due to changes in the OCu$_4$ tetrahedron. In the trigonal phase, the ∠Cu2–O3–Cu2 angle is ~114.7° (Fig. 3c). Symmetry reduction results in an increase in the number of symmetrically distinct Cu atoms and the splitting of intralayer Cu–O–Cu angles into three distinct values: 111.3°, 113.7° and 119.4°. Upon cooling, two of these angles decrease slightly (to 111.0° and 113.2°) while the third increases significantly to (121.1°). The fourth Cu atom, which caps the kagomé layer, exhibits a gradual inclination relative to the OCu$_4$ tetrahedron base, driven by the elongation of the Cu1–O2 bond (Fig. 4a). This results in a decrease in the angle between the Cu1–O5 bond and the tetrahedron base from ~86° to ~84°. The ∠Cu–Cu–Cu angles within the hexagonal rings of the kagomé layer also change notably (Fig. 4b). While one angle remains relatively constant (∠Cu2–Cu3–Cu2 ≈ 120°), the other two angle show significant variations: the ∠Cu3–Cu2–Cu2 decreases from ~104° to ~98°, and the ∠Cu2–Cu2–Cu3 increases from ~135° to ~140°. Thus, the ideal hexagonal ring of OCu$_4$ tetrahedra in the trigonal phase in the course of the phase transition experiences a ditrigonal distortion which is well-documented for silicate layers with identical structural topology.[30, 31]

The overall distortion of the $\{[O_2Cu_5](AsO_4)_2\}^0$ framework can be characterized by the relative shift of the [111] pyrochlore slabs. In the monoclinic phase, the unit cell is doubled along the $c$ axis, resulting in two such slabs within the unit cell related by an inversion center. Unlike the phosphate member of the averievite family,[12,14,28] these slabs do not exhibit significant shifts. However, due to the monoclinic symmetry, a slight shift of slabs in adjacent unit-cells persists along the $a$ axis. This shift remains relatively constant (0.15 Å at 280 K and 0.20 Å at 100 K).

**Thermodynamics**

Thermodynamic properties, i.e. magnetization $M$ and specific heat $C_p$, as well as both $ac$ and $dc$ magnetic susceptibility χ of (CsCl)Cu$_5$As$_2$O$_{10}$ were measured on a pressed pellet sample using various options of the "Quantum Design" Physical Properties Measurements System PPMS-9T.

Fig. 5a shows the $χ_{dc}(T)$ plot for (CsCl)Cu$_5$As$_2$O$_{10}$ measured at $μ_0H$ = 0.5 T in both field-cooled (FC) and zero field-cooled modes (ZFC) down to 2 K. This curve evidences the anomalies at both low and high temperatures. The high temperature anomaly enlarged in the inset to Fig. 5a is manifested in the divergence of the FC and ZFC curves in the range of $T_S$ = 305 – 315 K. It corresponds to the first order structural phase transition clearly seen in the diffraction data.

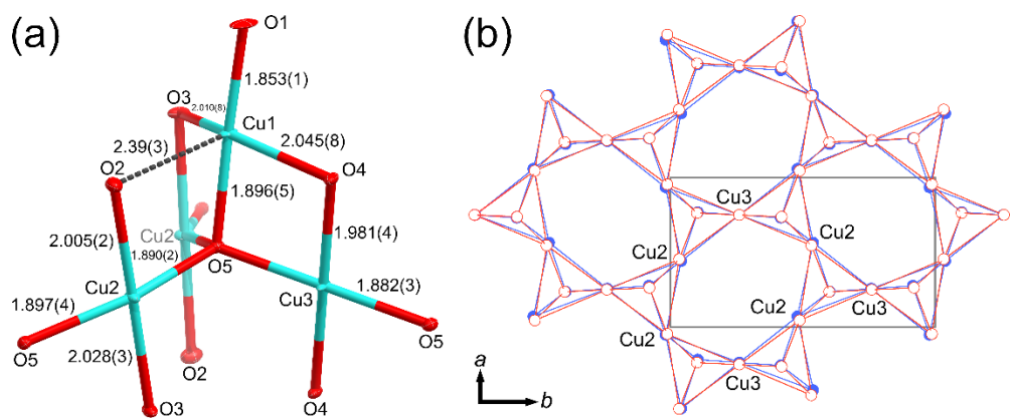

**Figure 4** OCu$_4$ tetrahedron (a) and the change of kagomé layer (b) with the temperature in the monoclinic (CsCl)Cu$_5$As$_2$O$_{10}$. Bond distances in (a) are averaged over the entire temperature range. Blue and red lines in (b) represent positions of Cu atoms and Cu–Cu distances at 100 and 280 K, respectively.

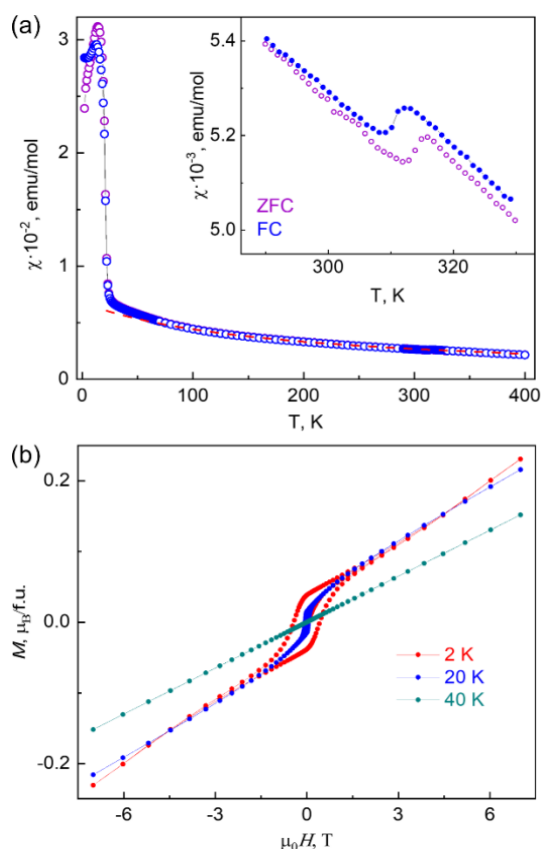

**Figure 5** (a) Temperature dependences of the magnetic susceptibility in (CsCl)Cu$_5$As$_2$O$_{10}$ taken in both FC and ZFC modes at $\mu_0 H$ = 0.5 T and (b) the field dependences of magnetization in (CsCl)Cu$_5$As$_2$O$_{10}$. The dashed line in (a) is the Curie–Weiss fit. The inset in (a) enlarges the anomaly at the structural phase transition.

In the range 140 − 285 K, i.e. below the structural phase transition, the FC curve was fitted by the Curie–Weiss law:

$$\chi = \chi_0 + \frac{C}{T - \Theta},$$

with temperature-independent term are $\chi_0$ = 1.1×10$^{-3}$ emu/mol, Weiss temperature $\Theta$ = − 139 K and Curie constant $C$ = 1.9 emu K/mol. This value of C leads to a reduced moment of Cu$^{2+}$ ions, nominally corresponding to a g-factor g = 2.

At $T_N$ = 21 K, both FC and ZFC $\chi_{dc}(T)$ curves evidence a sharp upturn followed by a peak at about 12 K. FC and ZFC curves separate below this temperature. The anomaly at $T_N$ marks the transition into a canted antiferromagnetic state, while the lower anomaly should be attributed to the domain processes. The values of Weiss and Neel temperatures allow estimating the frustration parameter $f = |\Theta|/T_N \sim 7$, which reflects low-dimensionality of the magnetic subsystem and frustration of the exchange interactions.

The presence of spontaneous magnetic moment in (CsCl)Cu$_5$As$_2$O$_{10}$ was confirmed in measurements of magnetization $M(H)$ at various temperatures, as shown in Fig. 5b. The hysteresis loops were seen at 2 and 20 K, while $M(H)$ at 40 K follows a linear trend. The saturation magnetization in (CsCl)Cu$_5$As$_2$O$_{10}$ $M_s$ is about 5.65 $\mu_B$/f.u. At residual magnetization $M_{res} \sim 0.03$ $\mu_B$/f.u., the canting angle of the two antiferromagnetic sublattices $\phi$ = 0.5°.

The temperature dependences of the ac-magnetic susceptibility in (CsCl)Cu$_5$As$_2$O$_{10}$ taken at the probing field 5 Oe in the frequency range 10$^{-3}$-10$^{-4}$ Hz are shown in Fig. 6a. These curves feature a sharp peak at $T_N$ = 21 K, whose position is independent of frequency. It excludes possible spin-glass effects. The frequency independence of the signal at $T < T_N$ indicates rather slow dynamics of the domain walls, which is reflected the dc-magnetic susceptibility only.

The temperature dependence of the specific heat $C_p$ in (CsCl)Cu$_5$As$_2$O$_{10}$ is shown in Fig. 6b. It evidences a λ-type anomaly at $T_N$, which signals a second order phase transition. The $C_p(T)$ curves taken at various magnetic fields at low temperatures are shown in the inset to Fig. 6b. These curves spread near $T_N$ and the anomaly is smeared out under magnetic fields.

**Nuclear magnetic resonance**

Nuclear magnetic resonance (NMR) spectra of $^{133}$Cs nuclei (nuclear spin $I$ = 7/2, gyromagnetic ratio γ = 5.62 MHz/T, quadrupole moment $Q$ = − 0.00355 barn) were recorded in a constant magnetic field of 5.5 T at various temperatures in the range of 7 − 300 K using an upgraded Bruker MSL spectrometer.[32] The standard Hahn spin echo method with a pulse sequence π/2 − π was used. $^{133}$Cs NMR spectra were obtained by Fourier transform of the second half of the spin

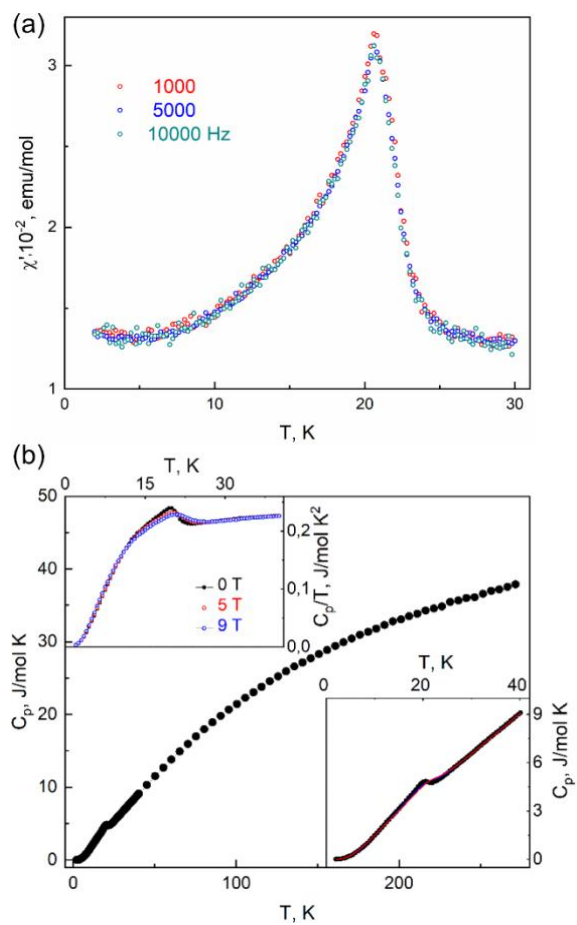

**Figure 6** (a) Temperature dependences of the ac-magnetic susceptibility at various frequencies in (CsCl)Cu$_5$As$_2$O$_{10}$ and (b) temperature dependence of the specific heat in (CsCl)Cu$_5$As$_2$O$_{10}$. The insets in (b) represent $C_p(T)$ curves at various magnetic fields.

echo for relatively narrow spectra above ~ 30 K and by a Fourier summation method at equidistant frequency points below ~ 30 K.[33,34] 10% aqueous solution of $CsNO_3$ was used as an NMR standard for $^{133}Cs$ nuclei. The nuclear spin-lattice relaxation rate, $1/T_1$, was measured by the saturation recovery method. The nuclear spin-spin relaxation rate, $1/T_2$, was measured by tracking the echo integral intensity with increasing the delay time τ between π/2 and π pulses.

$^{133}Cs$ NMR spectra of $(CsCl)Cu_5As_2O_{10}$ measured at various temperatures are shown in Fig. 7a. Above ~ 30 K, the line is narrow and symmetric with full width at half maximum (FWHM) less than 30 kHz, as shown in the inset to Fig. 7a. At decreasing temperature below ~ 30 K, the line exhibits steep broadening saturating at about 160 kHz at low temperatures. Since $^{133}Cs$ nuclei possesses extremely low quadrupole moment which is approximately 10 and 30 times less than that of $^7Li$ and $^{23}Na$ nuclei, respectively, one can assign the observed broadening exclusively to their magnetic nature excluding the quadrupole effects. The line broadening at low temperatures is accompanied by the appearance of a noticeable powder-like asymmetry of the $^{133}Cs$ NMR spectrum indicating an anisotropic distribution of local magnetic fields induced on the $^{133}Cs$ nuclei from magnetic copper ions in the magnetically ordered state.

At elevated temperatures, the variation of the $^{133}Cs$ NMR line shift K in $(CsCl)Cu_5As_2O_{10}$ is weak. It slowly decreases at lowering temperature, as shown in Fig. 7b. Below ~30 K it demonstrates an abrupt drop down to a saturated value of −150 kHz at low temperatures.

The temperature dependence of the $^{133}Cs$ nuclear spin-lattice relaxation rate, $1/T_1$, is shown in Fig. 7c. This parameter gradually increases with decreasing temperature until $T_N$, where $1/T_1$ exhibits a sharp kink. The spin-spin relaxation rate, $1/T_2$, of the $^{133}Cs$ nuclei is almost temperature independent above $T_N$, as shown in the inset in Fig. 7c. Below $T_N$, $1/T_2$ starts to grow and passes through a maximum at low temperatures.

The peculiarities observed by $^{133}Cs$ NMR spectroscopy in $(CsCl)Cu_5As_2O_{10}$ at low temperatures are in correspondence with results of thermodynamic measurements and manifest the formation of the magnetically ordered ground state. In general, the NMR shift consists of two main components:

$$K(T) = K_{spin}(T) + K_{orb},$$

where $K_{spin}(T)$ is the temperature dependent spin term and $K_{orb}$ is the orbital term, which is temperature independent. The spin component of the shift is proportional to the spin part of the dc magnetic susceptibility:

$$K_{spin}(T) \sim A_{hf} \cdot \chi_{spin}(T),$$

and hence provides a measure of $\chi_{spin}(T)$ which is insensitive to possible magnetic impurities. From Fig. 7b it is clear that $A_{hf}$ for $^{133}Cs$ nuclei in $(CsCl)Cu_5As_2O_{10}$ is negative and the observed $K(T)$ dependence reflects the presence of spontaneous magnetic moments in the ordered state of $(CsCl)Cu_5As_2O_{10}$. Similar behavior of $K(T)$ has been observed for $^{23}Na$ in ferromagnetic scutterudite $NaFe_4Sb_{12}$.[35] The broad hump in the spin-spin correlation rate, $1/T_2$, at low temperatures correlates with the anomaly in the $\chi_{dc}(T)$ curve at $T < T_N$, as well as with a minor curvature change on the $1/T_1$ dependence.

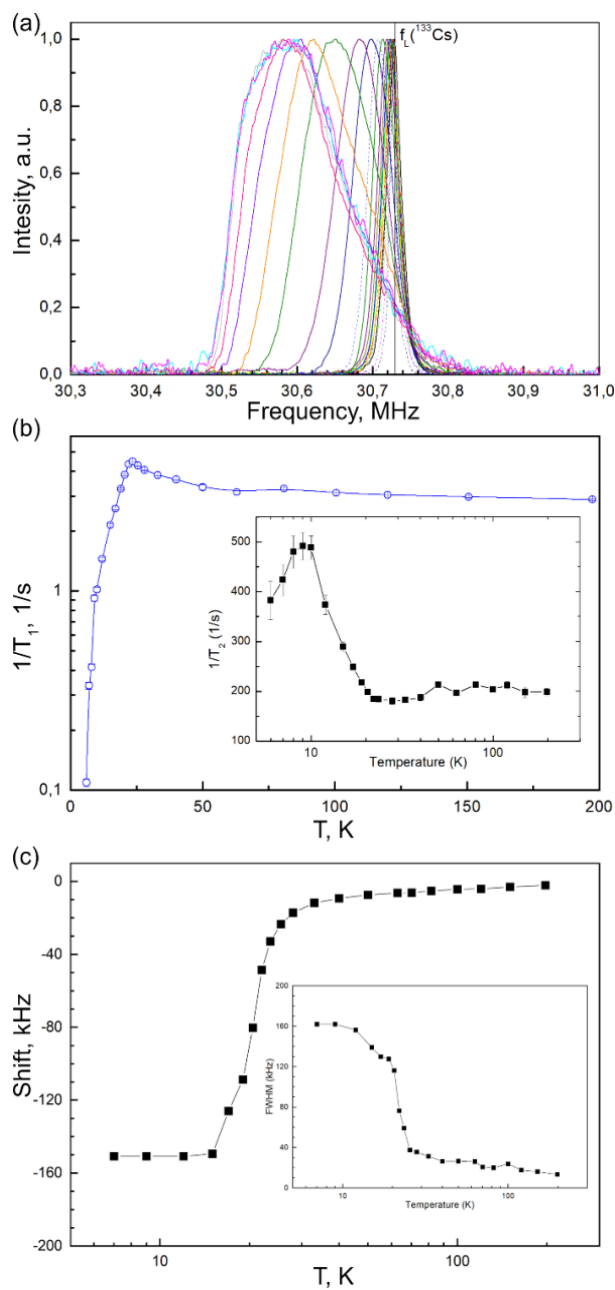

Figure 7 (a) $^{133}Cs$ NMR spectra of $(CsCl)Cu_5As_2O_{10}$ measured in constant magnetic field of 5.5 T at various temperatures in the range of 7 – 198 K (from left to right; vertical solid line denotes the Larmor frequency of $^{133}Cs$ nuclei measured in 10% aqueous solution of $CsNO_3$), (b) $^{133}Cs$ NMR line shift in $(CsCl)Cu_5As_2O_{10}$ as a function of temperature in relation to the Larmor frequency of $^{133}Cs$ nuclei in 10% aqueous solution of $CsNO_3$ (The inset represents the temperature dependence of the $^{133}Cs$ NMR FWHM) and (c) $^{133}Cs$ nuclear spin-lattice relaxation rate in $(CsCl)Cu_5As_2O_{10}$ as a function of temperature (inset: temperature dependence of the $^{133}Cs$ spin-spin relaxation rate).

**Magnetic Hamiltonian**

In order to work out the Heisenberg Hamiltonian parameters of As-averievite, we performed DFT calculations using the full potential local orbital (FPLO) basis set[36] in combination with the generalized gradient approximation exchange correlation functional.[37] We took into account strong electronic correlations on the $Cu^{2+}$ $3d$ orbitals using a DFT+U correction.[38] We use the DFT energy mapping approach to extract the parameters of the Heisenberg Hamiltonian

$$H = \sum_{i<j} J_{ij}\, S_i S_j,$$

where the sum runs over pairs of spins and every bond is counted only once. Energy mapping now implies that we calculate a large number of DFT+U energies for distinct spin configurations and fit them to the energies of Eq. (4). We created a four-fold supercell of As-averievite with 20 inequivalent $Cu^{2+}$ sites which allows us to extract 20 exchange interaction up to a Cu–Cu distance of about three times the nearest neighbor distance. With this approach, we have successfully investigated the magnetic properties of many copper-based magnets like the kagomé material barlowite $Cu_4(OH)_6FBr$,[39] the hyperkagomé system $PbCuTe_2O_6$,[40] the square kagomé antiferromagnet $Na_6Cu_7BiO_4(PO_4)_4Cl_3$.[41]

Fig. 8a shows the result of the DFT energy mapping for As-averievite. We show only four most prominent exchange interactions and report the smaller ones in the supplement. The network that is established by the two nearest neighbor couplings $J_1$=129(9) K and $J_2$=163(9) K is a pyrochlore slab as shown in Fig. 8b. A coupling that turns out to be significant is $J_6$=48(12) k which spans twice the nearest neighbor distance. Finally, the 2D layers of As-averievite are also significantly coupled in the third dimension; the most important of these 3D couplings is $J_8$=63(4) K. We can compare As-averievite to the DFT results of Dey and Botana.[16] Using our naming scheme of the couplings, they find the kagomé coupling to dominate in V-averievite at $J_1$ = 52K, $J_2$ = 228K, while the cap to kagomé coupling dominates for P-averievite at $J_1$ = 284K, $J_2$ = 235K.

The ideal nearest-neighbor (NN) pyrochlore Heisenberg antiferromagnet with all couplings equal features a flat band and the corresponding extensive degeneracy underpins its spin liquid behavior. In As-averievite, the NN couplings forming the tetrahedra split into two unequal ones, with three forming the basal triangles and three forming the cupolas. This leads to a lifting of the degeneracy and consequently the flat bands become dispersive albeit with a bandwidth which is much smaller compared to the total bandwidth. The inclusion of $J_6$ coupling (which in the ideal pyrochlore geometry would constitute only one set of those types of third nearest-neighbor interactions which are twice the nearest-neighbor) introduce a band gap with the upper bands. The Luttinger-Tisza minima are now located at incommensurate wave vectors, and classically this would lead to an energetic favoring of an incommensurate spiral order. The inclusion of further interlayer couplings $J_8$ are likely to strengthen the ordering tendencies. Indeed, within a classical Monte Carlo simulation

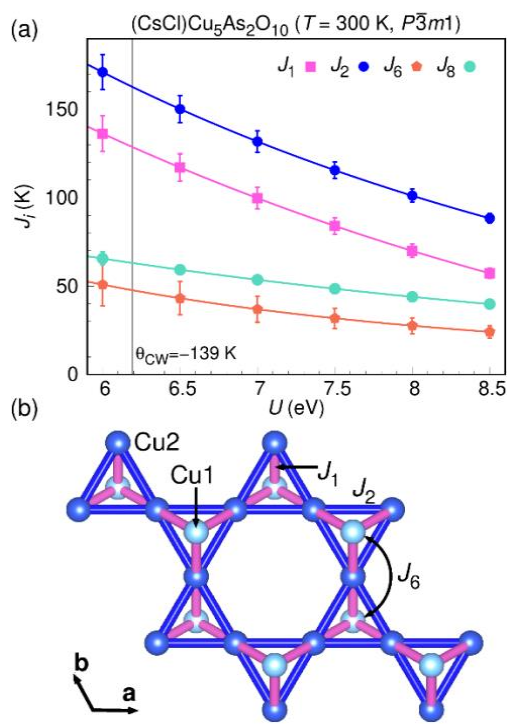

**Figure 8** (a) Exchange interactions of As-averievite obtained from density functional theory energy mapping. The four largest Heisenberg Hamiltonian parameters are shown as function of the strength of the on-site interaction $U$ in the GGA+U exchange correlation functional. The vertical line indicates the value of $U$ where the calculated exchange interactions match the experimental Curie-Weiss temperature. (b) Network of the three largest exchange interactions in $(CsCl)Cu_5As_2O_{10}$. They form a pyrochlore slab; the interlayer interaction $J_8$ is not shown.

of the Hamiltonian we find a sizable value of $T_c$ = 16.4 K as estimated from the behavior of the specific heat, as shown in Fig. S4 of SI. The calculated $T_c$ is lower compared to the experimentally observed value of 21 K, as would be expected from a simulation based on the more symmetric structure at room-temperature.[42]

We expect that the Hamiltonian based on the lower symmetry structure at 100 K temperature would yield a $T_c$ which is higher compared to the experimental value. We leave the determination of this Hamiltonian which would involve twice as many couplings, and its analysis to a future work. The fact that the experimental $T_c$ sits in between the $T_c$'s of these two structures indicates that the magnetic transition is accompanied by a structural distortion which accounts for the energy gain responsible for the onset of this transition.

## Conclusions

As shown above, the crystal structure of $(CsCl)Cu_5As_2O_{10}$ undergoes a structural first order phase transition from trigonal to monoclinic variety at ~310 K. As usual for the temperature-induced structural transformations, this transition is accompanied by the decrease of symmetry with the decreasing temperature. Information-theory analysis of structural complexity[43] shows that the structural information increases with the decreasing temperature from 2.760 bit/atom and 52.446 bit/cell for the trigonal phase to 3.406 bit/atom and 129.421 bit/cell for the monoclinic phase. This kind of behavior corresponds to the general trend of increasing complexity with decreasing temperature for structural phase transitions.[44]

The major driving force behind this transition is most probably associated with the ordering of $Cs^+$ cations in the cavities of the copper oxoarsenate framework. There are two low-occupied (0.14 and 0.36) sites in the trigonal phase, whereas there is only one fully occupied Cs site in the monoclinic phase. The Cs ordering is accompanied by the framework distortion with symmetry lowering induced by ditrigonal rotation of $OCu_4$ tetrahedra. Interestingly, both vanadate and phosphate analogs also undergo structural transitions, but their crystallographic nature differs significantly. The $(CsCl)Cu_5V_2O_{10}$ phase exhibits a structural phase transition at ~305 K from $P\bar{3}m1$ to $P\bar{3}$ with a doubling of the $a$ unit-cell parameter.[11,29] Another transition occurs at ~127 K, although the low-temperature crystal structure has yet to be resolved. A direct comparison of powder diffraction patterns of $(CsCl)Cu_5As_2O_{10}$ and $(CsCl)Cu_5V_2O_{10}$ indicates that $(CsCl)Cu_5V_2O_{10}$ does not adopt the same structural type as $(CsCl)Cu_5As_2O_{10}$. Meanwhile, the phosphate analogue, $(CsCl)Cu_5P_2O_{10}$, undergoes a first-order structural phase transition at ~224 K from centrosymmetric $P\bar{3}m1$ to non-centrosymmetric $P321$.[14,15] In this case, the kagomé layer loses its ideal hexagonal $p6mm$ symmetry, reducing down to $p3m1$ in a manner similar to $(CsCl)Cu_5As_2O_{10}$. However, this occurs without the relative shifting of pyrochlore slabs and while maintaining overall trigonal symmetry. As observed, the relationship between the volume of $T^{5+}O_4$ tetrahedra and the structural features is complex and remains poorly understood. Further investigation of potential solid-solution derivatives, their crystal structures, and the associated structural phase transitions is required to elucidate this relationship. Further interests also include investigating the effect of Cu-site substitution on magnetic properties and determining the magnetic structure of the title compound via neutron diffraction.[45]

Our thermodynamic studies, as well as the nuclear magnetic resonance measurements show that the quantum ground state of $(CsCl)Cu_5As_2O_{10}$ is a canted antiferromagnet formed at $T_N$ = 21 K, which is comparable with the ordering temperatures in $(CsCl)Cu_5P_2O_{10}$ (13.6 K) and $(CsCl)Cu_5V_2O_{10}$ (24 K). Notably, the ordering temperatures within the averievite family of compounds does not strongly depend on the exchange interactions of Cu ions within the kagome layers $J_2$ and with the decorating Cu ions $J_1$. Tentatively, the substitution of Cs by Rb in the averievite-type structure may increase the interlayer exchange coupling and the Neel temperature, correspondingly.

## Author contributions

Conceptualization – S.V.K., L.V.S., I.V.K.; investigation – I.V.K., M.V.L., I.E.L., A.V.T., S.V.Zh., L.V.S.; formal analysis – I.V.K., H.O.J., Y.I., L.V.S.; visualization – I.V.K., A.V.T., L.V.S.; writing – original draft – I.V.K., A.A.G., H.O.J., L.V.S., A.N.V.; writing – review & editing – S.V.K., A.N.V. All authors have given approval to the final version of the manuscript.

## Conflicts of interest

There are no conflicts to declare.

## Data availability

Crystallographic data for all compounds have been deposited at the ICSD under accession numbers 2480016-2480039 and can be obtained from [URL of data record, format https://doi.org/DOI].

## Acknowledgements

This work was supported by Russian Science Foundation, grant 24-17-00083. The X-ray diffraction and magnetization experimental data were carried out using equipment of the X-ray Diffraction Centre and Centre for Diagnostics of Functional Materials for Medicine, Pharmacology and Nanoelectronics of St. Petersburg State University. The research has been carried out within the framework of the scientific program of the National Centre of Physics and Mathematics under the project "Research in strong and superstrong magnetic fields". L.S. and A. V. acknowledge support by the Ministry of Science and Higher Education of the Russia within the framework of the Priority-2030 strategic academic leadership program at NUST MISIS. H.O.J. acknowledges support through JSPS KAKENHI Grant No. 24H01668. Part of the computation in this work has been done using the facilities of the Supercomputer Center, the Institute for Solid State Physics, the University of Tokyo. Y.I. acknowledges support from the ICTP through the Associates Programme, from the Simons Foundation through Grant No.~284558FY19, IIT Madras through the Institute of Eminence (IoE) program for establishing QuCenDiEM (Project No.~SP22231244CPETWOQCDHOC), and the International Centre for Theoretical Sciences (ICTS), Bengaluru for participating in the Discussion Meeting - Fractionalized Quantum Matter (code: ICTS/DMFQM2025/07). Y.I. also acknowledges the use of the computing resources at HPCE, IIT Madras. H.O.J. thanks IIT Madras for an IoE Visiting Faculty Fellow position during which this work was performed.